\documentclass[aps,onecolumn,amssymb]{revtex4}
\usepackage{epsfig,amssymb,amsmath}

\newcommand{\be}{\begin{equation}}
\newcommand{\ee}{\end{equation}}

\begin{document}
\title{Contact processes in crowded environments}
\author{S.-L.-Y. Xu$^1$ and J. M. Schwarz$^2$}
\affiliation{$^1$National Institutes of Health, Bethesda, MD 20892 and $^2$Physics Department, Syracuse University, Syracuse, NY 13244}
\date{\today}
\begin{abstract}
Periodically sheared colloids at low densities demonstrate a dynamical phase transition from an inactive to active phase as the strain amplitude is increased. The inactive phase consists of no collisions/contacts between particles in the steady state limit, while in the active phase collisions persist. To investigate this system at higher densities, we construct and study a conserved-particle-number contact process with novel three-body interactions, which are potentially more likely than two-body interactions at higher densities. For example, consider one active (diffusing) particle colliding with two inactive (non-diffusing) particles such that they become active, in addition to spontaneous inactivation. In mean-field, this system exhibits a continuous dynamical phase transition belonging to the conserved directed percolation universality class. Simulations on square lattices support the mean field result. In contrast, the three-body interaction requiring two active particles to activate one inactive particle exhibits a discontinuous transition. Finally, inspired by kinetically-constrained models of the glass transition, we investigate the ``caging effect'' at even higher particle densities to look for a second dynamical phase transition back to an inactive phase. Square lattice simulations suggest a continuous transition with a new set of exponents differing from conserved directed percolation, i.e. a new universality class for contact processes with conserved particle number. 

\end{abstract}

\maketitle

\section{Introduction}
Colloidal suspensions provide a platform to study interacting particle systems at lengthscales more easily accessible than the molecular scale. Successes of this platform range from evidence of caging in glassy colloidal suspensions~\cite{weeks} to progress towards colloidal self-assembly~\cite{assembly1,assembly2,assembly3} to the recent observation of a dynamical phase transition~\cite{pine,corte}. To observe this transition, a neutrally buoyant, non-Brownian colloidal suspension at low Reynolds number is periodically sheared using the standard Couette geometry.  Upon one shear cycle, some particles may collide with other particles.  See Figure 1.  Any collision/contact will displace the participating colloids from their typical cyclic shear profile and may trigger an avalanche of subsequent collisions. After some number of shearing cycles, the avalanche of collisions may, or may not, cease. In other words, the
system either eventually organizes itself into a state where every particle will not meet
others upon shearing or a state where collisions (however small the number) are unavoidable. The experiments demonstrate that one can tune between these two states by changing the strain amplitude. At small strain amplitudes, the collisions eventually cease.  At large strain amplitudes, they do not.

It has been argued that the existence of these two states---non-colliding and colliding---is a realization of an absorbing state phase transition~\cite{dickman,hinrich}, where the state with no collisions is the absorbing, or inactive, state~\cite{pine,corte}.  The collision phase, on the other hand, is the active phase since some fraction of particles are constantly being displaced from following their respective cyclic shear profile path.  This dynamical phase transition may indeed belong to the {\it conserved} directed percolation universality class~\cite{manna}.  

To understand this potential classification, we begin with {\it directed percolation}~\cite{harris,hinrich}. Directed percolation describes a system where ``activity'', i.e. collisions/contacts, may or may not be able to propagate through the system due to competing processes that either create or annihilate the activity. The creation of activity is dependent on the existence of preexisting activity nearby, hence, the directedness. The annihilation, on the other hand, is spontaneous. It has been argued that any continuous absorbing state transition of this kind with one absorbing state and no additional symmetries presumably falls under the directed percolation universality class~\cite{janssen,grassberger}. Despite the apparent ubiquitousness, experimental evidence for directed percolation emerged only recently in the electrohydrodynamic convection of liquid crystals~\cite{liquidcrystal}. 

We then ask what happens to the transition in directed percolation when particular symmetries are introduced?  For example, consider the conserved lattice gas (CLG) model~\cite{rossi}, where each site is assigned a value of zero (no particle) or one (one particle).  Particles repel each other such that particles with neighboring particles hop randomly to empty neighboring sites. These hopping particles are the active particles.  Isolated particles do not hop. Finally, double occupancy is not allowed. Particle number is conserved in this model such that an extra symmetry is introduced. In addition, there exist many configurations with no active particles.  Simulations of this CLG model as the density is increased indicate a transition from a state with no active particles to a state with active particles in the steady state limit. It turns out that the measured critical exponents differ from directed percolation indicating a universality class distinct from directed percolation~\cite{rossi,pastor,vespignani,lubeck1,lubeck2,lee}. 

The colloids under cyclic shear experiment also has the property that particle number is conserved. In addition, it contains many possible absorbing states and local activation via collisions. So, at some level, the experiment is a version of the CLG model where the strain amplitude tunes the activation rate as opposed to the particle density. However, what if more details of the experiment are incorporated into the modeling?  Just how robust is the conserved directed percolation universality class? More specifically, if one modifies the rules of the conserved lattice gas model, does the system exhibit a similar transition? For example, if the diffusion of the active particles is not isotropic but rather dependent on the direction of the colliding particles, would the CDP exponents be robust?  This modification to an off-lattice version of the CLG model measures CDP exponents at/near the absorbing state transition~\cite{menon}. Other lattice models containing local, stochastic activation, many absorbing states, and an additional symmetry also exhibit CDP exponents~\cite{mansur,lubeck3}. A more recent periodically sheared colloidal experiment with elliptical particles exhibits CDP exponents as well~\cite{fibers}.

\begin{figure}[t]
\begin{center}
\epsfig{figure=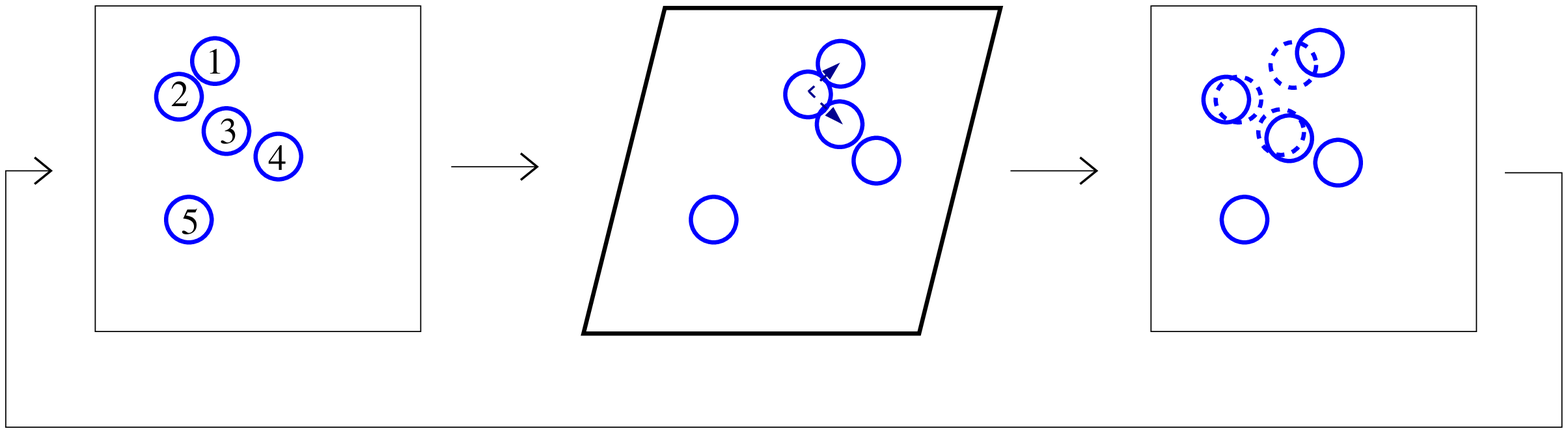,width=5.0in}
\caption{A two-dimensional schematic of the three-dimensional experiment presented in Ref. 6 during the one shear cycle. Particles 1, 2, and 3 are active particles since they are displaced from their initial positions (drawn in dashed lines on the far right box), while particles 4 and 5 are inactive particles.  Particle 4 may become active on the next shear cycle, while particle 1 may become inactive.}
\end{center}
\label{fig:schematic}
\end{figure}

Now, the initial experiment was conducted at low packing fraction such that most collisions are presumably two-body collisions. We pose the following question: What happens to the dynamics of the colloids as the packing fraction is increased at fixed strain amplitude?  Also, what happens to the dynamics when the packing fraction is increased towards the colloidal glass transition? Increasing the packing fraction increases the possibility of three or more particle collisions, thereby justifying the study of higher-order collision, or contact, processes. Of course, from a renormalization group perspective as applied to equilibrium transitions, higher-order interactions are less relevant and so one may not expect the scaling near the absorbing state phase transition to differ from the lowest-order interaction case unless the rate of two-body collisions was tuned to zero.  While this may be difficult in an experiment, it is not the case for simulations. On other hand, nonequilibrium phase transitions are not necessarily equivalent to equilibrium ones. 

Therefore, we analytically and numerically study higher-order contact processes with active and inactive particles. The activation and inactivation rules are local, the total particle density is conserved, and there are many absorbing states. In addition to studying the role of three or more particle collisions on the absorbing state phase transition, we study the effect higher density has on the diffusion of the active particles, or lack thereof. If the active particles are prevented from diffusing due to surrounding particles via caging~\cite{weeks}, then the active particles are rendered inactive. In the highly dense limit, the system will ultimately become inactive due to this additional density dependent inactivation. We will incorporate such an effect into the model and study the properties of this presumably new absorbing state phase transition. This new absorbing state transition could potentially describe a colloidal glass transition under cyclic shear and merges the two previously independent concepts of contact processes and the glass transition.

The paper is organized as follows: Section II establishes the model we study and presents our analytical results in mean field, Section III relays our numerical results in two-dimensions, and Section IV summarizes and discusses the implications of our results. 
 
\section{Model and mean field analysis}

Consider a system with two types of particles: (1) particle type A denotes the active members and (2) particle type B denotes the inactive ones. The inactive particles are "inactive" because they do not diffuse, while the active particles diffuse. We denote the density of each particle type as $\rho_A$ and $\rho_B$ respectively with the total density $\rho=\rho_A+\rho_B$. We consider the following reaction processes:
\begin{eqnarray}
A&\stackrel{r+q(\rho_A,\rho_B)}{\rightarrow}&B\nonumber\\
cA+dB&\stackrel{f(c,d)}{\rightarrow}&(c+d)A\nonumber,
\end{eqnarray}
where $r$ denotes the density independent inactivation rate, $q(\rho_A,\rho_B)$ denotes the density dependent inactivation rate, and $f(c,d)$ denotes the activation rate that changes with the number of particles involved. These rules conserve total particle number no matter which order of interaction (values of $c$ and $d$) is present. The density dependent inactivation rate will depend on the local particle density and models the caging effect. Specific implementation will be addressed later. 

As with directed percolation, it is useful to build and study the equations of motion for this set of reactions in mean field (averaging out the spatial degrees of freedom) and in the absence of any noise. Mean field theory is simple and sheds light on the nature of the phase transition---is it continuous or discontinuous, for example?  Typically, a transition that is continuous in mean field is also continuous in finite-dimensions, though there are exceptions to the rule~\cite{cardy}.  

Let us first analyze the case previously studied as a model for conserved directed percolation, i.e. $c=1$, $d=1$, $f(1,1)=s$, and $q=0$. Then, the reaction rules are
\begin{eqnarray}
A&\stackrel{r}{\rightarrow}&B\nonumber\\
A+B&\stackrel{s}{\rightarrow}&A+A,
\end{eqnarray}
where $s$ is the activation rate upon contact between an active particle and an inactive one, and $r$ is the spontaneous inactivation rate of active particles regardless of its neighborhood. In mean field, this model yields a dynamical equation for the density of active particles of
\begin{equation}
\frac{d\rho_A}{dt}=-r\rho_A+s\rho_A\rho_B.
\end{equation} 
Since the overall density is conserved, $\rho=\rho_A+\rho_B$, once we determine $\rho_A$, we also know $\rho_B$, i.e.
\begin{equation}
\frac{d\rho_A}{dt}=-r\rho_A+s\rho\rho_A-s\rho_A^2.
\end{equation} 
In steady state, 
\begin{equation}
\rho_{As}(\rho-\frac{r}{s}-\rho_{As})=0,
\end{equation}
and we observe two solutions for the steady state active particle density, $\rho_{As}$,
\begin{equation}
\rho_{As}=
\begin{cases}
0, & \rho<\rho_c=\frac{r}{s}, \\
\rho-\rho_c, & \rho\ge\rho_c,
\end{cases}
\end{equation}
predicting a continuous phase transition with linear scaling,
$\rho_A=(\rho-\rho_c)^\beta$ for $\rho\ge\rho_c$ with order parameter exponent
$\beta=1$. See Fig. 2 for a mean-field phase diagram. 

The mean-field correlation length exponent, $\nu_{\perp}=1/2$, and the mean field dynamic exponent, $z=2$, can be can be obtained by including the diffusive nature of the active particles into the dynamical equation for $\rho_A$ to arrive at
\begin{equation}
 \frac{d\rho_A}{dt}=D_A\nabla^2\rho_A-r\rho_A+s\rho\rho_A-s\rho_A^2.
\end{equation} 
After rescaling the equation with $\rho_A\rightarrow \lambda \rho_A$, $x\rightarrow \lambda^{-\nu_{\perp}}x$, and $t\rightarrow \lambda^{-\nu_{||}}t$, scale invariance is observed when $s\rho-r=0$, $\nu_{\perp}=1/2$, and $\nu_{||}=1$ such that $z=\nu_{||}/\nu_{\perp}=2$. The addition of the noise term demonstrates a critical dimension of four.   

\begin{figure}[t]
\begin{center}
\epsfig{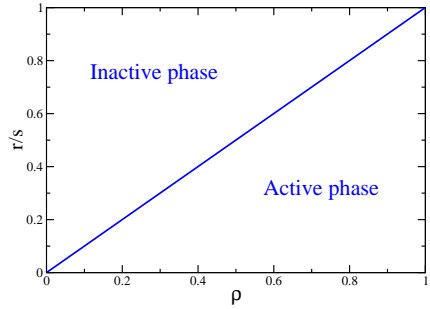}
\caption{Mean-field phase diagram for the $A+B$ activation process.}
\end{center}
\label{fig:ABmfphasedia}
\end{figure}

As the particle density is increased, it is reasonable to consider three-or-more body interactions because an active particle can collide with more than one inactive particles in its path, for example. One of the next-lowest-order contact processes is 
\begin{eqnarray}
A&\stackrel{r}{\rightarrow}&B\nonumber\\
A+2B&\stackrel{u}{\rightarrow}&A+2A.
\end{eqnarray}
As above, in mean field we have
\begin{eqnarray}
\frac{d\rho_{As}}{dt}=-r\rho_{As}+2u\rho_{As}\rho_{Bs}^2&=&0\nonumber\\
\rho_{As}((\rho-\rho_{As})^2-\frac{r}{2u})&=&0,
\end{eqnarray}
with the solution
\begin{equation}
\rho_{As}=
\begin{cases}
0, & \rho<\rho_c=\sqrt\frac{r}{2u},\\
\rho-\rho_c, & \rho\ge\rho_c.
\end{cases}
\end{equation}
We obtain the same type of continuous transition as before with a shift in the transition location, however.  This result is not unexpected since the inactive particles are, in a sense, enslaved to the active ones.  They do not contain their own inherent dynamics.  The rescaling analysis of the dynamical equation with diffusion added back in also results in $\nu_{\perp}=1/2$ and $z=2$.  

If we have a mixture of the two lowest-order interactions (and this is presumably closer to what happens in reality), then the reaction rules are
\begin{eqnarray}
A&\stackrel{r}{\rightarrow}&B\nonumber\\
A+B&\stackrel{s}{\rightarrow}&A+A\\
A+2B&\stackrel{u}{\rightarrow}&A+2A\nonumber.
\end{eqnarray}
These rules correspond to a steady state equation of
\begin{eqnarray}
\frac{d\rho_{As}}{dt}=-r\rho_{As}+s\rho_{As}\rho_{Bs}+2u\rho_{As}\rho_{Bs}^2,
\end{eqnarray}
with the steady state solution
\begin{equation}
\rho_{As}=
\begin{cases}
0, & \rho<\rho_c=\frac{\sqrt{s^2+8ru}-s}{4u} \\
\rho-\rho_c, & \rho\ge\rho_c.
\end{cases}
\end{equation}
The solution does not appear that intuitive. Let us examine the condition under which the transition occurs, that is $0<\rho_c<1$. Since both the numerator and the denominator are positive for any positive values of the transition rates $r$, $s$, $u$, the condition $\rho_c>0$ is always satisfied to guarantee an absorbing phase. On the other hand, the upper bound $\rho_c<1$ can be shown to be equivalent to $r<s+2u$. In a mean field system, close to the absorbing transition, all active particles are exposed to a continuum background of inactive particles, and the net production of active particles is proportional to $s+2u-r$. In summary, a combined lowest-order and next-lowest-order contact process exhibits, again, a continuous absorbing phase transition with order parameter exponent $\beta=1$. 

There is another type of three-body contact process in the form of
\begin{eqnarray}
A&\stackrel{r}{\rightarrow}&B\nonumber\\
2A+B&\stackrel{v}{\rightarrow}&2A+A.
\end{eqnarray}
This new activation process corresponds to the activation of an inactive particle occuring when two active particles impact it at the same time. The mean field dynamical equation of these processes is 
\begin{eqnarray}
\frac{d\rho_{As}}{dt}=-r\rho_{As}+v\rho_{As}^2\rho_{Bs}.
\end{eqnarray}
The absorbing state solution of $\rho_{As}=0$ is still valid. Moreover, using $\rho_{Bs}=\rho-\rho_{As}$, we arrive at
\begin{equation}
s\rho_{As}^2-v\rho\rho_{As}+r=0,
\end{equation}
whose solution is 
\begin{equation}
\rho_{As}=\frac{\rho}{2}\pm\frac{1}{2}\sqrt{\rho^2-\frac{4r}{v}}.
\end{equation}
This solution is physical only when $\rho\ge\rho_c=\sqrt{4r/v}$, and at the
critical point $\rho=\rho_c$, $\rho_{As}=\rho/2=\sqrt{r/v}\neq0$. We have thus
generated a discontinuous absorbing phase transition at the mean field level. This
activation process is very different from the one driven by a single active
particle, which is continuous. See Figure 3 for the mean-field phase diagram for this process. 

\begin{figure}[t]
\begin{center}
\epsfig{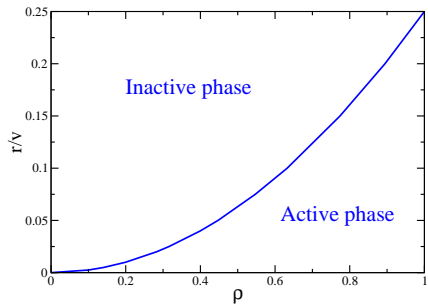}
\caption{Mean field phase diagram for the $2A+B$ activation process.}
\end{center}
\label{fig:2ABmfphasedia}
\end{figure}

As the density of particles in the system is increased even further, there will eventually be less effective volume for the active particles to diffuse in, thus, they become inactive. In a lattice model, we model this effect with an active particle becoming an inactive one when all of its neighboring sites are occupied by active or inactive particles. Of course, neighboring active particles will eventually diffuse away, though not before inhibiting the motion of the active particle. We consider a lattice with coordination number $Z$. Then, in mean field,  the caging effect is equivalent to adding another inactivation process with rate, $w\rho^Z$. If we consider the lowest-order contact process with this new caging effect, the reaction rules are now
\begin{eqnarray}
A&\stackrel{r+w\rho^Z}{\rightarrow}&B\nonumber\\
A+B&\stackrel{s}{\rightarrow}&A+A,
\end{eqnarray}
with corresponding mean field equation of motion:
\begin{equation}
\frac{d\rho_{As}}{dt}=-(r+w\rho^Z)\rho_{As}+s\rho_{As}\rho_{Bs},
\end{equation}
whose steady state solutions are
\begin{equation}
\rho_{As}=
\begin{cases}
0, \\
\rho-\frac{\rho^Z}{s}-\frac{r}{s},
\end{cases}
\end{equation}
for $w=1$.

In general, the absorbing transitions occur at the roots of equation: $-\rho^Z/s+\rho-r/s=0$. We denote the roots as $\rho_c$, and within a vicinity of $\rho_c$ with $\rho=\rho_c+\delta\rho$ where $\delta\rho\ll\rho_c$, we have
\begin{eqnarray}
\rho_{As}&=&\rho-\frac{\rho^Z}{s}-\frac{r}{s}\nonumber\\
&=&\rho_c+\delta\rho-\frac{(\rho_c+\delta\rho)^Z}{s}-\frac{r}{s}\nonumber\\
&=&\rho_c+\delta\rho-\frac{1}{s}\rho_c^Z(1+\frac{\delta\rho}{\rho_c})^Z-\frac{r}{s}\nonumber\\
&\approx&\rho_c+\delta\rho-\frac{\rho_c^Z}{s}(1+Z\frac{\delta\rho}{\rho_c})-\frac{r}{s}\nonumber\\
&=&\delta\rho-\frac{Z}{s}\rho_c^{Z-1}\delta\rho\nonumber\\
&=&(1-\frac{Z}{s}\rho_c^{Z-1})\delta\rho\nonumber\\
&\sim&(\rho-\rho_c)^\beta,
\end{eqnarray}
where $\beta=1$. For any value of $Z$, as long as the absorbing transition occurs (in a certain volume of $\{Z,r,s\}$ phase space), the transition is continuous with order parameter exponent $\beta=1$. As an example, for a 1D lattice with $Z=2$, the active phase solution is
\begin{equation}
\rho_{As}=-\frac{\rho^2}{s}+\rho-\frac{r}{s},
\end{equation}
and we see an active phase $\rho_{As}>0$ when $\rho_{c-}<\rho<\rho_{c+}$ where $\rho_{c\pm}=\frac{1}{2}(s\pm\sqrt{s^2-4r})$. The condition for an active phase to exist is $s^2-4r\ge0$. The transition at $\rho_{c-}$ is the usual absorbing phase transition at low density, and the reentrant absorbing phase at higher density represents the second transition due to caging. See Figure 4. 

\begin{figure}[t]
\begin{center}
\epsfig{figure=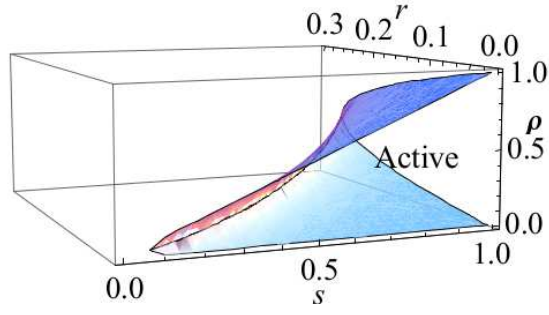,width=3.0in}
\caption{Mean field phase diagram including the caging effect for $Z=2$.}
\end{center}
\label{fig:cagingpd}
\end{figure}

As an alternative model of the caging effect, one may think that if an active particle is caged by other active particles, the active particle might actually be unblocked by its neighbors and then it is able to move. Based on this argument, an active particle turns inactive by caging only if its neighboring sites are all occupied by inactive particles, i.e.
\begin{eqnarray}
A&\stackrel{r+w'\rho_B^Z}{\rightarrow}&B\nonumber\\
A+B&\stackrel{s}{\rightarrow}&A+A.
\end{eqnarray}
The corresponding mean-field dynamical equation is
\begin{equation}
\frac{d\rho_{As}}{dt}=-(r+\rho_{Bs}^Z)\rho_{As}+s\rho_{As}\rho_{Bs}
\end{equation}
with $w'=1$. There is still an absorbing state solution of $\rho_{As}=0$. The active steady state solution corresponds to the solution of $s(\rho-\rho_{As})-r-(\rho-\rho_{As})^Z=0$. If one assumes that the transition occurs at $\rho_{As}=0$, then the critical value of $\rho=\rho_c$ is the root of equation $s\rho-r-\rho^Z=0$, and it is clear that $\rho_{As}=\rho-\rho_c$, which indicates, again, a continuous transition with order parameter exponent $\beta=1$. 

In sum, when considering higher-order activation processes where the number of inactive particles involved is increased beyond one, at the mean field level, the transition appears to be the same as the lowest-order activation contact process.  This is because the inactive particles are enslaved to the active ones. The additional density-dependent inactivation contact process to model caging does not alter the order parameter exponent in mean field either. Only when additional active particles are incorporated into the activation process, does even the nature of the transition change from continuous to discontinuous. Now we will investigate this system beyond mean field. 

\section{Simulations}

We begin by elaborating on the simulation protocol and then present our simulation results. 

\subsection{Algorithm}

Consider a square lattice of linear size $L$ with periodic boundary conditions.  For initialization, the particles are distributed on sites according to a preset density $\rho$, which is equivalent to the occupation probability of a site. The initial fraction of active particles is denoted by $a$. Each site is occupied by at most one particle at any time, i.e. no double occupancy. 

At every time step,\\

(1) Each active particle hops to one of its empty neighboring sites with equal probability, i.e. diffusion. Active particles must hop as long as they have at least one empty neighboring site. If no empty neighbor is available, the active particle does not hop and becomes inactive (caging effect). \\

(2) Each active particle can activate one nearest inactive neighbor at random with probability $\bar{s}$, where the bar denotes the rate being converted to a probability working in units of the maximum density set to unity and the simulation time step. As for three-body interactions, one active particle can activate two neighboring inactive particles with probability $\bar{u}$, two active particles can activate one commonly neighboring inactive particle with probability $\bar{v}$.\\


(3) Each active particle becomes inactive independent of its neighborhood with probability $\bar{r}$. \\

The simulation is carried out until steady state is reached defined by the number of active particles on average not changing with time. From the steady state mean-field analysis, the ratio of $\bar{r}$ to $\bar{s}$, for example, alters the competition between activation and inactivation and, therefore, determines the usual absorbing state transition point. In lattice simulations, however, there are corrections to the $\bar{r}/\bar{s}$ ratio determining the critical point. Finally, sample averaging typically ranges from $64,000$ for the smallest system size to $2,000$ for the largest system size for the $A+2B$ and the $2A+B$ activation processes, and $100,000$ for the smallest system size and $1,000$ for the largest system size for the caging inactivation process occuring at higher densities. 

\subsection{Results}

We are primarily interested in how the system behaves as a function of the total density, $\rho$, since changes in the total density drive the system from the absorbing phase to the active phase and vice-versa at high densities. We will look for such transitions in (1) the $c=1,d=2$ model with no caging, (2) the $c=2,d=1$ model with no caging, and (3) the $c=1,d=1$ model with caging. 

Should the absorbing state transition appear to continuous, one would expect a series of scaling relations similar to the ones measured in numerical simulations of conserved directed percolation models~\cite{rossi,pastor,vespignani,lubeck1,lubeck2,lee}. In particular, the density of active particles at the critical point would decay as a power-law in time, or $\rho_A\sim t^{-\theta}$. Also, the correlation time would scale by the distance to transition, or $\tau\sim|\rho-\rho_c|^{\nu_\parallel}$. Thus, one expects a scaling function in the form of
\begin{equation}
\rho_A(t)=t^{-\theta}F(t/\tau)=t^{-\theta}F(t/|\rho-\rho_c|^{\nu_\parallel}),
\end{equation}
where $F(x)$ is a universal off-critical scaling function. Moreover, the steady state density should follow $\rho_{As}\sim(\rho-\rho_c)^\beta$ in the long time limit, $t\gg\tau$. Hence, the scaling relation $\beta=\theta\nu_\parallel$ should hold similar. 
	
As for other scaling checks in the case of a continuous absorbing state transition, the correlation length near the critical point scales with the linear system size $L$ as $\xi\sim |\rho-\rho_c|^{\nu_\perp}\sim L$ and, thus, $|\rho-\rho_c|\sim L^{-1/\nu_\perp}$. The scaling of $\rho_A$ with both time $t$ and system size $L$ can be described as
\begin{equation}
\rho_A(t)=t^{-\theta}G(t/L^z),
\end{equation}
where $z=\nu_\parallel/\nu_\perp$ is the dynamic exponent and $G(x)$ is another 
universal finite-size scaling function.  

\subsubsection{$A+2B$ activation mechanism}

For a given set of parameters, we must first hone in on the transition. As we found in mean field, if the inactivation probability is much higher than the activation probability, then the initial inactive-to-active transition will not occur.  Working with following probabilities: $a=1/2$, $\bar{r}=1/4$, and $\bar{u}=1/2$ (and all other probabilities zero), we find the inactive-to-active transition around $\rho=0.45$.  Our simulation results for $\rho_A(t)$ at different total densities near $\rho=0.45$ are plotted in Fig. 5. As the total density increases, the system transitions from an absorbing phase to an active phase, i.e. the steady states value of $\rho_A$ going from zero to non-zero.

Not only do we expect this particular transition to be continuous, we speculate it to be in the same universality class as conserved directed percolation. This speculation is consistent with the property 
that the inactive particles are ``enslaved'' to the active ones.  In other
words, they cannot move around without the help of the active ones so that changing the number of inactive particles activated does not change
the universality class (as it may with other higher-order interactions becoming relevant in equilibrium phase transitions).  In other words, the dynamics underlying the A+2B activation process is no different from the A+B activation. On the other hand, changing the number of initially activated particles, i.e. 2A+B, should change the universality class since the dynamics is now modified and the activation becomes cooperative. We will return to this speculation later. 

To check our first speculation---that the $c=1,d=2$ transition is in the same universality class as conserved directed percolation (CDP)---we check for numerical consistency with this universality class. In other words, we use the exponents obtained from prior work for CDP to collapse the data. The most precise values of $\theta$, $\nu_{||}$, and $z$ for CDP are $\theta=0.410$, $\nu_{||}=1.544$, and $z=1.53$~\cite{lee}. Using the first two values in the offcritical scaling with $\rho_c$ as a fitting parameter, we are able to obtain a good scaling collapse with $\rho_c=0.4502(1)$.  In other words, the data is rather consistent with the corresponding exponents of conserved directed percolation universality class~\cite{lee}. 

\begin{figure}[t]
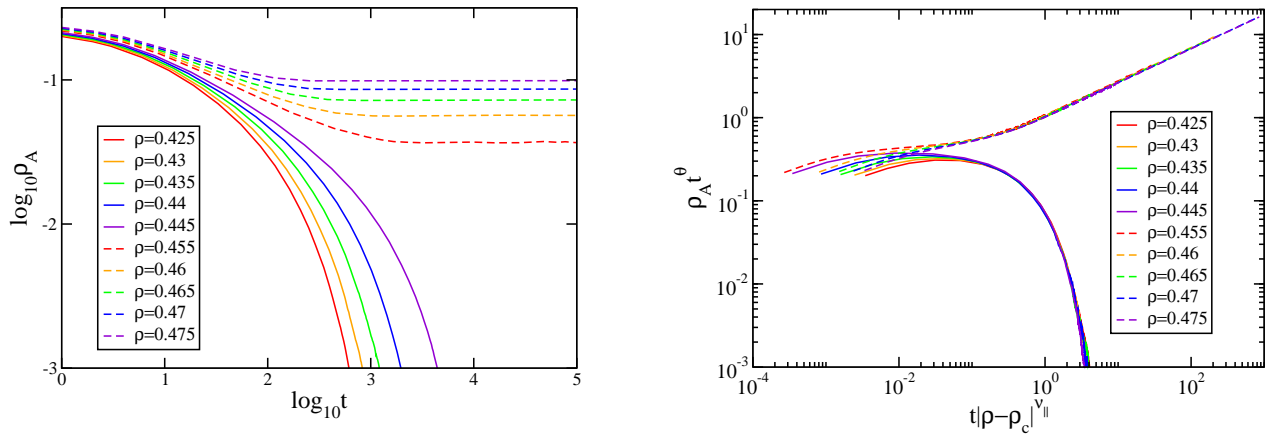

\vspace{0.5cm}
\begin{center}
\epsfig{figure=A2B.offcritical.L512.unscaled.final.eps,width=3.0in}
\hspace{1cm}
\raisebox{-0.2cm}{\epsfig{figure=A2B.offcritical.L512.scaled.final.eps,width=3.1in}}
\caption{Left: Plot of the $\log_{10}(\rho_A)$ versus $\log_{10}(t)$ for the $A+2B$ contact process at different total occupation probabilities with $L=512$. Right: Log-log plot of the off-critical scaling collapse, $\rho_At^\theta$
  vs. $t|\rho-\rho_c|^{\nu_\parallel}$ with $\rho_c=0.4502(1)$ and using the CDP values of $\theta$ and $\nu_\parallel$.}
\end{center}
\label{fig:A2Boffcritical}
\end{figure}

\begin{figure}[t]
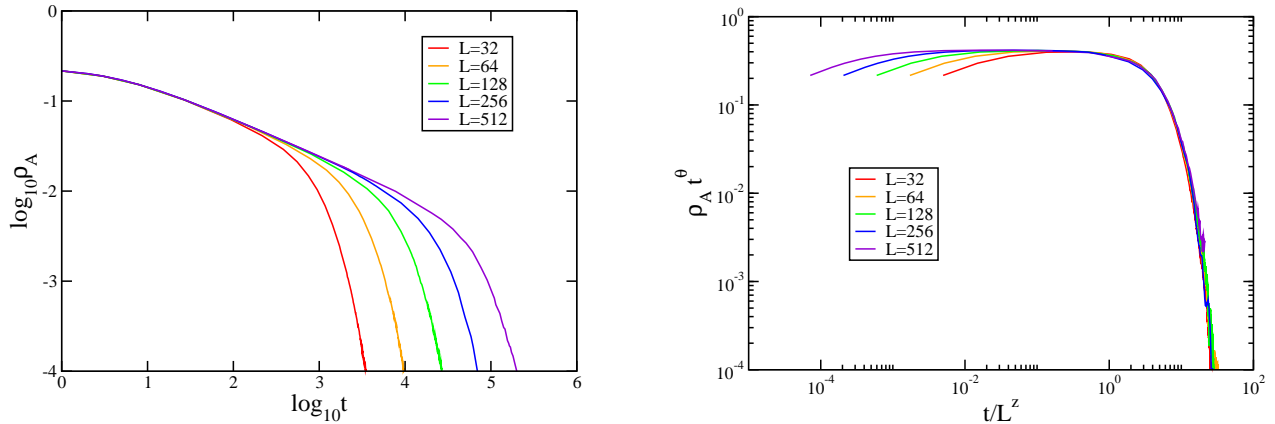

\vspace{0.5cm}
\begin{center}
\epsfig{figure=A2B.finite.size.p0.4502.unscaled.final.eps,width=3.0in}
\hspace{1cm}
\raisebox{-0.0cm}{\epsfig{figure=A2B.finite.size.p0.4502.scaled.final.eps,width=3.1in}}
\caption{Left: Plot of the $\log_{10}(\rho_A)$ versus $\log_{10}(t)$ at the critical density, $\rho_c$, for different system sizes. Right: Log-log plot the for finite-size scaling collapse at the critical point, or $\rho_At^\theta$ vs. $t/L^z$, again, using the CDP values of $\theta$ and $z$.}
\end{center}
\label{fig:A2Bfinitesize}
\end{figure}

To further our claim that the A+2B activation process is in the same universality class as CDP, we obtain $\rho_A(t)$ at the previously obtained $\rho_c$ for different system sizes. Due to the finite-size fluctuations, the system will eventually reach the absorbing state in some finite time at the critical density. This is explicitly shown in Fig. 6. Power-law decays at early times are clear with a universal power. At late times finite-size fluctuations drive the fall-off of the order parameter.  If we use the CDP values of $\theta$ and $z$, we, again find that the data are very well
collapsed according to the above scaling relation with CDP exponents. Note there is no free parameter for this finite-size scaling collapse. 

Given these numerical observations, we conclude that the A+2B activation process falls in the conserved directed percolation universality class.

\subsubsection{$2A+B$ activation mechanism}
What about the $2A+B$ activation process? In mean field we find that a discontinuous transition occurs, which is very different from the continuous transition found in CDP models.  Figure 7 is a plot of $\rho_A(t)$ for different $\rho$s using $a=1/2$, $\bar{r}=1/10$, and $\bar{v}=9/10$. We observe that these curves look qualititatively
different from the $A+2B$ case. The standard scaling collapse is not
possible here with the data suggesting a discontinuous transition. From these results, we conclude that the $2A+B$ activation is very different from the $A+2B$ activation process and leave such cooperative activations for future work and turn to the inactivation via caging process.

\begin{figure}[t]
\begin{center}
\vspace{0.5cm}
\epsfig{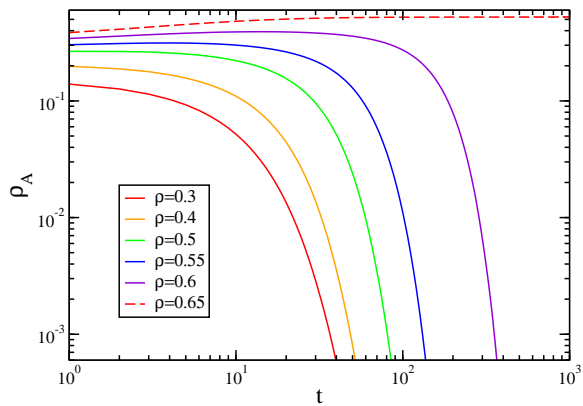}
\caption{Plot of the $\log_{10}(\rho_A)$ versus $\log_{10}(t)$ for the $2A+B$ contact process at different total occupation probabilities with $L=512$. }
\end{center}
\label{fig:2ABoffcritical}
\end{figure}

\subsubsection{Caging inactivation mechanism} 

The caging mechanism in which an active particle becomes inactive if no empty neighboring site is present at any given time step does not affect the $c=d=1$ absorbing phase transition nor the $c=1$, $d=2$ transition at ``low'' densities. Close to the first dynamical transition, the density of particles is low enough such that there is almost always at least one available empty site for an active particle to diffuse. However, as we increase the particle density the caging effect becomes more significant. We should observe a second absorbing phase transition at high densities back to an inactive state where active particles turn inactive due to local geometrical constraints. In other words, the caging mechanism brings in a local density dependent inactivation rate in addition to the spontaneous inactivation invoked in CDP. We choose to work with the total density of neighboring particles inactivating active particles, as opposed the density of inactive neighboring particles particularly since the parameter space for observing the caging transition with the latter rule is smaller, i.e. a more restrictive constraint. 

According to the mean-field analysis, this caging transition is a continuous
transition with order parameter exponent $\beta=1$. In other words, it should be same type as the
first absorbing state phase transition at low densities. Should the transition be continuous in finite-dimensions as well, then the scaling
functions $F(x)$ and $G(x)$ constructed above should apply and data should show
the same type of scaling behavior. However, we expect a different set of exponents here since the mechanism of inactivation is changed from a density independent one to a density dependent one such that the competition between activation and inactivation is rather different from CDP.  After obtaining $\rho_A(t)$ for different total densities with $a=1$, $\bar{r}=1/4$, $\bar{s}=1/2$, and $\bar{w}=1$, a scaling collapse with $\theta$, $\nu_\parallel$, and $\rho_c$ as parameters is found for $\theta=0.32(1)$, $\nu_\parallel=1.32(2)$, and $\rho_c=0.8838(1)$. See Fig. 8. The two exponents are rather different from the CDP class exponents, suggesting a new universality class. For example, $\theta=0.410$ in the latter case and $\theta=0.32(1)$ in the former. In addition, here, $\nu_\parallel=1.32(2)$, which is to be contrasted with the value of $1.544$ for CDP.

\begin{figure}[t]
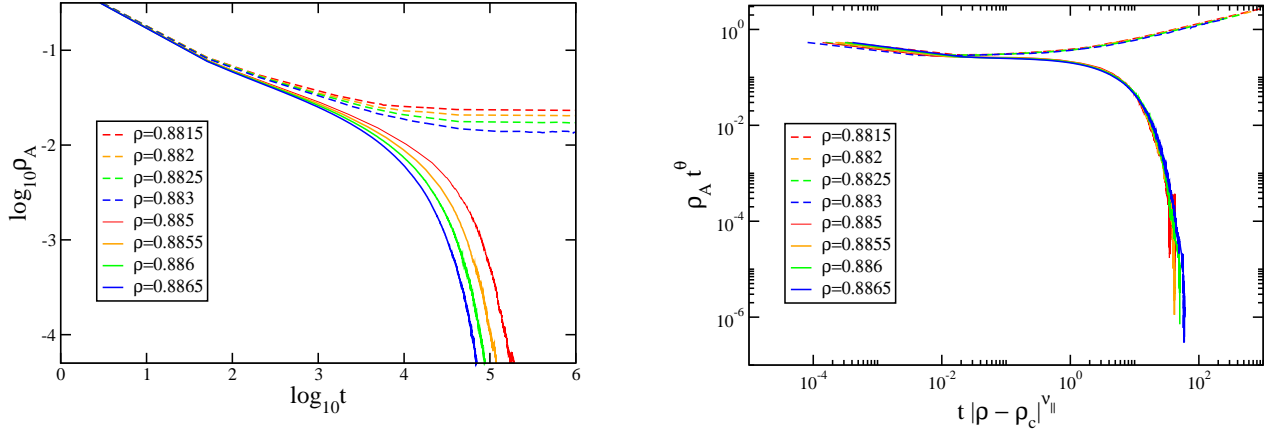

\vspace{0.5cm}
\begin{center}
\epsfig{figure=caging.offcritical.L512.unscaled.final.eps,width=3.0in}
\hspace{1cm}
\raisebox{-0.2cm}{\epsfig{figure=caging.offcritical.L512.scaled.final.eps,width=3.1in}}
\caption{Left: Plot of the $\log_{10}(\rho_A)$ versus $\log_{10}(t)$ at different total occupation probabilities for the caging transition with $L=512$. Right: Log-log plot of the off-critical scaling collapse, $\rho_At^\theta$
  vs. $t|\rho-\rho_c|^{\nu_\parallel}$ with $\rho_c=0.8838(1)$, $\theta=0.32(1)$ and $\nu_\parallel=1.32(2)$.}
\end{center}
\label{fig:cagingoffcritical}
\end{figure}

We also investigate finite-size scaling at the caging transition. Collapse of the data is obtained using $\theta=0.32(1)$ to find that $z=2.11(2)$. See Fig. 9. Again, this $z$ is rather different from the conserved directed percolation of $z=1.53$ such that we have presumably discovered a new absorbing state universality class.

\begin{figure}[t]
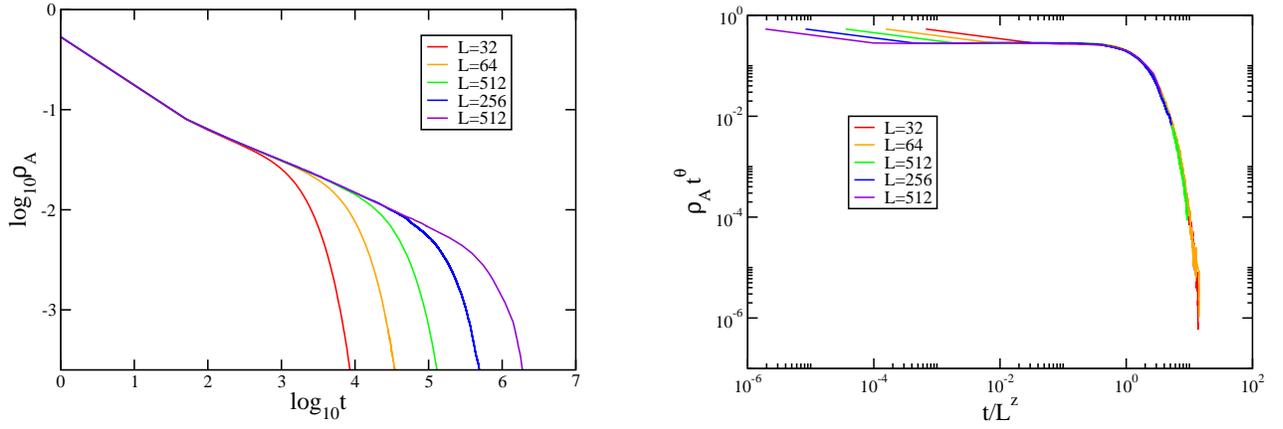

\vspace{0.25cm}
\begin{center}
\epsfig{figure=caging.finite.size.p0.8838.unscaled.final.eps,width=3.0in}
\hspace{1cm}
\raisebox{-0.0cm}{\epsfig{figure=caging.finite.size.p0.8838.scaled.final.eps,width=3.1in}}
\caption{Left: Plot of the $\log_{10}(\rho_A)$ versus $\log_{10}(t)$ at the critical density, $\rho_c$, for different system sizes. Right: Log-log plot for finite size scaling at the critical point, or $\rho_At^\theta$ vs. $t/L^z$ with $z=2.11(2)$.}
\end{center}
\label{fig:cagingfinitesize}
\end{figure}

To look for more evidence of a
continuous transition, we analyze the
relation between system size and the magnitude of the jump in the order
parameter, the steady state density of active particles, at the transition due to finite-size
fluctuations.
Beginning with $\rho_A(t)=t^{-\theta}G(t/L^z)$, using $z\theta=\beta/\nu_\perp$ yields  
\begin{equation}
\rho_A(t)=L^{-\beta/\nu_\perp}H(t/L^z).
\end{equation}
For $t>>L^z$ in the active phase, $\rho_A(t)$ approaches a constant such that $\rho_A(t)\rightarrow \rho_{sat}\sim L^{-\beta/\nu_{\perp}}$ as the transition is approached from the active phase. Since $\rho_A(t)$ eventually vanishes in any finite-sized system, to measure $\beta/\nu_{\perp}$ directly, the usual procedure has been to include only the surviving samples in the averaging to obtain the size of the jump in the order parameter~\cite{rossi}. See Fig. 10 for such data where $\rho_A^*$ denotes the average over surviving samples only. The slope in the inset of Fig. 10 corresponds to $-\beta/\nu_\perp=-1.1$. From earlier measurements, however, the scaling relations $\beta=\theta\nu_\parallel$ and $\nu_\perp=\nu_\parallel/z$ imply that $\beta=0.42$ and $\nu_\perp=0.62$. Then, $\beta/\nu_\perp=0.69$, which is not consistent with our measurement suggesting $\beta/\nu_\perp=1.1$. A similar broken scaling relation has been reported in the conserved lattice model~\cite{rossi}. Lee and Lee~\cite{lee} point out that the $z$ obtained from the all-sample averaging data was employed in the surviving-sample averaging. Instead, they argue that if the above scaling relation is tested using the all-sample averaging, consistency with $\beta=\nu_{||}\theta$ and $z=\nu{_||}/\nu_{\perp}$ is observed. We observe the same consistency when using the Lee and Lee~\cite{lee} scaling approach.  Lee and Lee~\cite{lee} then went on to investigate $\rho_A^*(t)$ to find that the saturation time of $\rho_A^*(t)$ is different from time scale for finite-size effects to affect the system. In other words, there are two different time scales. This finding appears to be particular to conserved contact processes, i.e., it is not found in directed percolation.  They argue that only all-sample averaging data should be used to obtain the critical exponents~\cite{lee}. 

\begin{figure}[t]
\begin{center}
\epsfig{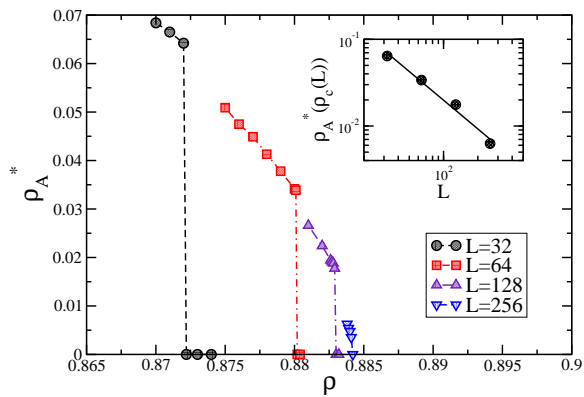}
\caption{Surviving sample averages for $\rho_A^*(\rho)$ for different system sizes. Inset: Log-log plot of the magnitude of the jump in $\rho_c^*$ at $\rho_c(L)$ versus $L$. The line drawn through the data has a slope of $-1.1$.}
\end{center}
\label{fig:jamjump}
\end{figure}

In sum, study of the caging mechanism results in the potential discovery of a new universality class for the active-inactive transition at high densities.  See Table 1 for a comparison of the CDP exponents with the newly obtained caging transition exponents. 

\begin{table}[ht]
\setlength{\tabcolsep}{12pt}
\centering
\begin{tabular}{|c| c |c |}
\hline
& CDP & Caging\\
\hline
$\theta$ & $0.410$ & $0.32(1)$\\ \hline
$\nu_{\parallel}$ & $1.544$ & 1.32(2)\\ \hline
$z$ & $1.53$ & $2.11(2)$ \\ 
\hline
\end{tabular}
\caption{Comparison of conserved directed percolation (CDP) exponents obtained in Ref. 19 with the caging transition exponents obtained here.} 
\end{table}

\section{Discussion}

Inspired by the observation of an absorbing state phase transition in a colloidal system under cyclic shear, we introduce and study higher-order contact processes to explore the possibility of new absorbing state phase transitions at higher densities than previously explored in experiments. Therefore, we extend the two-body collision/contact process framework to include more than two-body collisions, which are more likely at high densities. 

We find that a simple three-body collision of one active (diffusing $A$) particle activating two inactive (nondiffusing $B$) particles does not change the nature of the absorbing phase transition found at lower densities due to two-body collisions. More precisely, the exponents obtained for the two-body collision transition, namely the conserved directed percolation exponents, can also be used to the collapse the two-dimensional data for the $A+2B$ activation process~\cite{corte,menon}.  This is due to the fact that the inactive particles are enslaved to the active particles such that the inactive particles simply form a continuous background that an
active particle can always access. In other words, the difference between
activating one inactive particle ($A+B$) and activating two inactive
particles ($A+2B$) is effectively a change in the activation rate and not in the
fundamental mechanism of activation.

On the other hand, further investigation of the three-body $2A+B$ 
activation process would be promising in terms of
looking for a new transition of a very different nature as indicated by both mean-field calculations and two-dimensional simulations. The $2A+B$ process requires that the active particles must be clustered into groups of at least two to generate
more active particles to stay in the active phase. Moreover, isolated active particles would be eliminated very
fast since they cannot reproduce. Spontaneous inactivation combined with a ``tougher'' activation condition make it unlikely that the transition is continuous with a small fraction of active particles just above the transition. The possibility of a discontinuous transition in the cooperative activation model may be 
comparable to a recent modification of the asymmetric exclusion process where a particle can hop to its right only if its left neighbor is occupied to arrive at very different features from the usual asymmetric exclusion process~\cite{gabel}. Finally, it is not clear whether the $2A+B$ activation process would be more likely than the $A+2B$ activation process in the cyclically sheared colloid experiment at higher densities.  It would also be interesting to explore how this new discontinuous phase transition relates to other discontinuous transitions observed in several contact process models~\cite{lee2,silva}.

At even higher densities, we expect a caging-type inactivation mechanism to occur in cyclically sheared colloids because active particles become trapped by neighboring colloids and are unable to diffuse. In other words, they become inactive.   
We incorporate such a mechanism into the usual conserved directed percolation model. We numerically observe in two-dimensions, a ``caging'' transition at high densities where the active state becomes inactive. This transition appears to be in a new universality class distinct from CDP. We anticipate that this transition could be observed in cyclically sheared colloids at high density. 

We expect this new active-to-inactive transition at high densities to effectively be a colloidal glass transition under cyclic shear, which may be different from the colloidal glass transition in the absence of cyclic shear~\cite{glass}. The cyclic shear provides a simple, but dynamic, reference state allowing for a precise definition of active and inactive particles.  The construction of active and inactive particles is different from the kinetically constrained models of the glass transition~\cite{FA,kob}. In the kinetically constrained models, rules constrain the movements of the particles, and the onset of a percolating cluster of particles that cannot move constitute the glass transition. Such rules always obey detailed balance. With such rules, it has been difficult to observe a transition in finite-dimensions~\cite{tbf}, though more complex rules allow for a transition~\cite{jeng,tb}.  Kinetically constrained models where the rules do not obey detailed balance have recently been introduced by studying the glass transition in the presence of a biasing field (as opposed to changing temperature)~\cite{elmatad}. This biasing field induces a new finite-temperature glass transition.  The biasing field is somewhat analogous to external forcing, i.e. cyclic shear, since the dynamics need not obey detailed balance. 

While periodically sheared colloidal systems give us insight into nonequilibrium phase transitions at the micron scale, periodically sheared granular systems give us insight into nonequilibrium phase transitions at the millimeter scale. Experiments on granular systems at densities near jamming indicate the presence of dynamical hetereogeneities, which could be interpreted as indirect evidence of the caging phenomenon~\cite{candelier}. In light of the dynamical transition found in the colloidal system, experimentalists have recently looked for this possibility in a granular system~\cite{slotterback}. When probing the mean-squared displacement of the particles and the fraction of broken links, there is evidence of different behavior between a strain of 20 degrees and 40 degrees. However, only when measuring the fraction of grains in the largest cluster, which is a measure of the global topology, is there some evidence for a phase transition. It could be that this transition is related to the dynamical transition observed in the colloidal system, though it is a transition in the topology and not the spatial organization, since reversibility in particle displacements was not observed in the dense, granular system (due to collisions).  More recently, controlled simulations were able to access this regime~\cite{schreck}. However, in addition to a loop-reversible regime where despite collisions, the particles completed closed loops such that they returned to their initial positions.

In closing, this work merges for the first time the fields of contact processes and their associated dynamical phase transitions with the colloidal glass transition via the introduction of caging into a contact process model. The result is two-fold: (1) The discovery of a new universality class in the realm of dynamical phase transitions and (2) a model for the onset of glassiness in a periodically sheared colloidal system. As for the former, the Grassberger/Janssen conjecture states inactive-active transitions with a positive one-component order parameter, short-range interactions, and no additional symmetries or quenched disorder should fall into the directed percolation universality class~\cite{janssen,grassberger}. Particle number conservation falls outside this conjecture to yield conserved directed percolation.  But just how many possible universality classes exist within the umbrella of conserved particle number? The data presented here suggests that it may be more than one. As for the onset of glassiness, the field has been plagued with a lack of microscopic understanding for years such that a concrete microscopic model for the onset of glassiness in the presence of cyclic shear may serve as a basis for modeling the glass transition in general.

JMS kindly acknowledges discussions early on in the project with Bismayan Chakrabarti on conserved particle versions of $k$-core percolation and funding support from NSF-DMR-CAREER-0645373.

\end{document}